\documentclass[preprint]{elsarticle}
\usepackage{amssymb}

\def\tect{$^{130}$Te }
\def\tectn{$^{130}$Te}

\def\udt{$^{238}$U }

\def\tld{$^{208}$Tl }

\def\coss{$^{60}$Co }

\def\amnun{$\vert\langle m_{\nu} \rangle\vert$}

\def\BBz{$\beta\beta(0\nu)$~}
\def\BBzn{$\beta\beta(0\nu)$}

\def\BB{$\beta\beta$~}

\def\ca{$\sim$}

\def\conteggi{counts/kg\,keV\,yr\,}
\def\gammaconteggi{$\gamma$/s\,cm$^2$\,}
\def\muconteggi{$\mu$/s\,cm$^2$\,}
\def\neuconteggi{n/s\,cm$^2$\,}

\def\pom{$\pm$ }

\def\teod{TeO$_2$~}

\def\be{\begin{equation}}
\def\ee{\end{equation}}

\def\ciccio{5$\times$5$\times$5 cm$^3$ }

\begin{document}
\begin{frontmatter}

\title{Monte Carlo evaluation of the external gamma, neutron and muon induced background sources in the CUORE experiment}

\author[rm]{F.Bellini}
\author[LNGS]{C.Bucci}
\author[MIB]{S.Capelli\corref{COR}}\ead{silvia.capelli@mib.infn.it}
\author[MIB]{O.Cremonesi}
\author[MIB]{L.Gironi}
\author[MIB]{M.Martinez}
\author[MIB]{M.Pavan}
\author[LNGS]{C.Tomei}
\author[rm]{M.Vignati}

\cortext[COR]{Corresponding author: tel: +39 02 6448 2417; fax: +39 02 6448 2463}

\address[rm]{Universit\`a di Roma, Dipartimento di Fisica and INFN, I-00185 Roma, Italy}

\address[LNGS]{INFN, Laboratori Nazionali del Gran Sasso, I-67100 L'Aquila, Italy}

\address[MIB]{Universit\`a di Milano Bicocca, Dipartimento di Fisica and INFN, I-20126 Milano, Italy}

\begin{abstract}
CUORE is a 1 ton scale cryogenic experiment aiming at the measurement of the Majorana mass of the electron neutrino. The detector is an array of 988 \teod bolometers used for a calorimetric detection of the two electrons emitted in the \BBz of \tectn. The sensitivity of the experiment to the lowest Majorana mass is determined by the rate of background events that can mimic a $\beta\beta(0\nu)$.
In this paper we investigate the contribution of external sources i.e. environmental gammas, neutrons and cosmic ray muons to the CUORE background and show that the shielding setup designed for CUORE guarantees a reduction of this external background down to a level $< 10^{-2}$ \conteggi at the Q-value, as required by the physical goal of the experiment.
\end{abstract}


\end{frontmatter}


\section{Introduction}
The sensitivity of an experiment looking for very rare events such as \BBz can be spoiled by background counts in the region of interest (ROI). The most dangerous background sources for such experiments are due to cosmogenic activation and intrinsic contamination of the detector and setup materials, and to environmental sources in the experimental hall.   
In a deep underground laboratory the environmental background can be ascribed to three different sources: gammas, neutrons and cosmic ray muons. Gammas and neutrons are due both to the natural radioactivity of the rock constituting the laboratory walls and to muon interactions in the rock, in the materials surrounding the detector and in the detector itself. 
The reduction of the gamma and neutron background in an underground experiment is generally achieved by means of passive shields while a muon identification system is used  to tag and reject  muon induced events.
Shields and vetoes are designed to match the specific experimental features, mainly by considering the kind of physical results the experiment aims to achieve and considering the various radioactive sources that could spoil the experimental sensitivity.

In this paper we discuss the impact of the above mentioned external sources on the sensitivity of the CUORE experiment. In section~\ref{sec:CUORE} we describe the experiment and its physical goals, in section~\ref{sec:SOURCES} we list the available information about the environmental background in the experimental site, while the following sections are devoted to the description of the Monte Carlo simulation code used for the background estimates and of the results obtained in terms of contributions to the CUORE background from the different external sources.

\section{The CUORE Project}
\label{sec:CUORE}

The nature of neutrino mass is one of the frontier problems of fundamental physics. Neutrinoless Double Beta Decay (\BBzn) is a powerful tool to investigate the nature of the neutrino and in the Majorana hypothesis the mass hierarchy ~\cite{09Petcov} and possible extensions of the Standard Model.  The \BBz is a rare spontaneous nuclear transition where a nucleus (A,Z) decays into an (A,Z+2) nucleus with the emission of two electrons and no neutrinos, resulting in a monochromatic peak in the sum energy spectrum of the electrons. The \BBz decay is possible only if the neutrino is a Majorana massive particle. The transition width is proportional to the square of \amnun, the effective Majorana mass of the electron neutrino. From the \BBz half-life it is therefore possible to infer the correct mass hierarchy and the absolute mass scale of neutrino:

\begin{equation}
(T_{1/2}^{0\nu})^{-1}=G^{0\nu}(E_0 , Z) \bigg| \frac{\langle m_{\nu} \rangle}{m_{e}} \bigg|^2 \big| M^{0\nu}_{F}-(g_{A}/g_{V})^2 M^{0\nu}_{GT} \big|^2
\end{equation}

\noindent where $G^{0\nu}$ is a phase space factor including the couplings, $M^{0\nu}_{F}$ and $M^{0\nu}_{GT}$ are the Fermi and Gamow-Teller nuclear matrix elements, respectively, and $g_{A}$ and $g_{V}$ are the relative axial vector and vector weak coupling constants, respectively.

The CUORE (Cryogenic Underground Observatory for Rare Events) experiment ~\cite{arnaboldi04} aims at searching for \BBz of \tect with a background $<10^{-2}$ \conteggi and an effective  Majorana mass sensitivity of a few tens of meV, reaching the so called inverse hierarchy region of the neutrino mass spectrum ~\cite{09Petcov}.
It is presently under construction in the Hall A of the Gran Sasso Underground Laboratory (LNGS) at an average depth of \ca3600 m.w.e. ~\cite{LVD}. 

\subsection{The Detector}\label{sec:DetStruct}

The CUORE detector is a system of 988 cryogenic bolometers. The array is composed by 19 vertical towers cooled  by means of a dilution ($^3$He/$^4$He) refrigerator to a temperature of \ca10\,mK. A tower consists of 13 layers of 4 bolometers each. 
The bolometer consists of a \ciccio \teod cubic crystal, an NTD Ge thermistor, and a heater (a Si resistor doped with As) glued with Araldit Rapid Epoxy onto the crystal surface ~\cite{arnaboldi04}. 
The dielectric TeO$_2$ crystal has a  very low heat capacity and a very large ($27\%$ in mass) natural isotopic abundance of the \BBz candidate \tect. Each cube has a weight of 750 g, so the  detector is a granular calorimeter of total mass 741\,kg, corresponding to \ca200\,kg of \tect.

The  cryogenic detecting technique was  proposed for the first time by Simon~\cite{Simon} to study nuclear phenomena and  then  suggested to search for rare events by Fiorini and Niinikoski~\cite{Fiorini}. 

The typical performance of such detectors has been measured in previous experiments: MIDBD~\cite{MiDBD} the first experiment operating a large array of bolometers to search for \BBz, CUORICINO~\cite{Cuoricino}, a single CUORE  tower running for five years at LNGS and other tests performed with CUORE prototype detectors. They have shown an average energy FWHM resolution of 1-2~keV at 10~keV and 5~keV at 3~MeV and an extremely slow time development of the signal pulse. The 90-10\% average rise time was evaluated to be of about 50 ms, and the 90-30\% decay time  of about 100 ms, with a low tail in its distribution that extends even above one second. Typical energy thresholds for double beta decay searches are kept between 50 and 300 keV. 

\begin{figure}
\begin{center}
\includegraphics[width=0.5\linewidth]{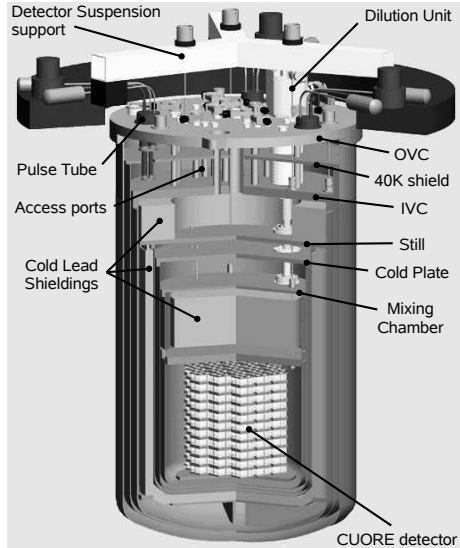}
\end{center}
\caption{3D view of the cryogenic apparatus.}
\label{cryostat}
\end{figure}

The CUORE cryogenic apparatus is shown in Fig.~\ref{cryostat}. The cryostat is made of six nested vessels and its base temperature without heat loads is expected to be as low as 6\,mK.
Three lead shields are used to protect the detector from environmental radioactivity and from contaminations in the building materials. A 25\,cm thick lead layer outside the OVC shields the detector from radiations coming from the bottom and from the sides. An equivalent shielding against radiation coming from the top has to be placed inside the cryostat, just above the detector. This is a 30\,cm thick lead disk with a diameter of about 90\,cm. Just below it copper disks totaling an additional 8\,cm shields are placed. A lead ring-shaped shield placed on the Steel flange closes the gap between the lead disk above the detector and the outer room temperature shield. 
An additional shielding of detector's sides and bottom is provided by a 6\,cm lead layer just outside the Steel shield.  
 
Outside the external lead shield an 18 cm thick polyethylene layer will be added in order to thermalize environmental neutrons that will then be absorbed by a 2 cm layer of H$_3$BO$_3$ powder contained in the hollow space between the lead and the polyethylene itself. On the top and on the bottom of the setup, due to construction difficulties, the neutron shield will be made of a single 20 cm layer made of polyethylene with 5\% in boron. 

\subsection{Physics goals}

Operated as a single particle detector, the cubic bolometer will be able to detect a \BBz  occurring in its bulk by recording the two electrons emitted in the decay. The two electrons (and the recoiling nucleus) deposit their whole energy in the crystal where the decay takes place with a probability of 86.4\% (computed with a Monte Carlo  simulation). In this case, the bolometer records a monochromatic line at the \BBz transition energy (2527.517 \pom 0.013 keV for \tect \cite{Avignone09,Scielzo09}), while the other bolometers of the array are not involved in the decay.
Background events that can mimic a \BBz decay and hence spoil the sensitivity of the experiment are those that correspond to an energy deposition similar to that of \BBzn. In CUORE, a large fraction of those events  can be rejected simply requiring the signal to be a single-hit event as it is expected for a \BBz decay. The reduction factor depends on the actual source kind and position as shown in Tab.~\ref{tab:backgroundDBD}. 
Monte Carlo simulations show that this reduction can range from a factor 2 for bulk contaminations of the shields and crystal surface contaminations to two orders of magnitude for the cosmic muon contribution.
The probability of a spurious time coincidence between a background event in one crystal of the CUORE array and a \BBz event in another crystal is P$_{\rm{coinc}}$ = $\Delta$T$ \cdot $N$_{\rm{DET}}\cdot$\emph{f}, being $\Delta T$=50 ms a reasonable value for the coincidence window, corresponding to the signal rise time,  N$_{\rm{DET}}$=988 the number of detectors and \emph{f} the average rate.
Assuming for \emph{f} the value measured on CUORICINO detectors of 5 mHz we obtain the conservative estimate P$_{\rm{coinc}}\sim 25\%$.
This probability can be reduced by means of a near channel selection, without loosing in terms of background rejection capability. Taking into account only the 18 closest crystals the probability of spurious coincidences is reduced to 0.45\%.

The experimental sensitivity is conventionally parametrized in terms of the number of \BB emitters (N$_{\beta\beta}$), the detection efficiency ($\epsilon$), the live-time ($T$), the energy resolution ($\Gamma$) and the background counting rate ($B$): 

\begin{equation}
S_{0\nu\beta\beta}= \rm{ln}2\times N_{\beta\beta}\times\epsilon\times\sqrt{\frac{ M \times T}{\Gamma\times B}}
\end{equation} 

The background counting rate of CUORE is obviously the ultimate parameter that will determine the \BBz sensitivity. CUORICINO has measured a background level of 0.18 \conteggi with a 40.7~kg array of CUORE-like detectors. A smaller array consisting of eight \ciccio detectors assembled in a two planes CUORE-like tower, after a dedicated surface treatment of the copper structures and of the crystals, has been recently operated  obtaining a background rate of 0.06 \conteggi. CUORE points toward an even better background level ($<10^{-2}$ \conteggi) exploiting its granularity, the specific design of  detector, cryostat and shields and the careful selection and preparation of materials.

\section{External background sources}
\label{sec:SOURCES}

As already stated, the possibility to study rare events such as \BBz is strongly influenced by the background in the ROI. There are various sources that give rise to spurious counts in this region such as environmental gamma radioactivity, cosmic rays, neutrons, radon and contamination of materials which detectors and their shielding are made of. The evaluation of the background due to cosmogenic activation and intrinsic contamination of the detector components is described in ~\cite{pavan08}. Radon emanation in the experimental hall is not an issue for CUORE since all the gamma lines of the$^{222}Rn$ have energies below the \tect \BBz Q-value. In this paper we focus on the contribution coming from external background sources: environmental gamma radioactivity, cosmic rays (muons) and neutrons.

\subsection{Gammas} \label{sec:SOURCES_gammas}
\begin{figure}
\begin{center}
\includegraphics[width=0.8\textwidth]{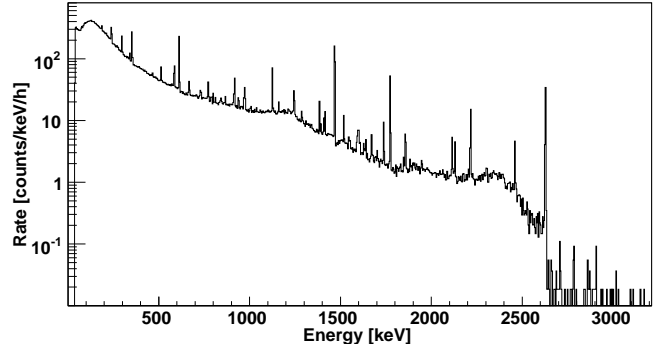}
\end{center}
\caption{Gamma ray spectrum measured at LNGS (Hall A) with a small Ge diode.}
\label{fig:gammaLNGS}
\end{figure}

The gamma spectrum due to natural rock radioactivity at LNGS extends up to 3 MeV while above that energy the main contribution to the gamma background comes from photons produced by neutron and muon interactions in the rocks. Due to their high energies and lower flux (estimated to be $10^5$ times smaller than the flux of the 2.6 MeV gamma-ray of \tld in the rock ~\cite{07Pandola})  these photons cannot be measured with a small Ge diode.   
In Fig.~\ref{fig:gammaLNGS} the gamma-ray spectrum, as measured by a small Ge diode in the Hall A of LNGS, is shown.  The reconstructed flux is reported in Tab.~\ref{tab:gammaflux}  \cite{Bucci09}. The integral gamma ray flux below 3 MeV has been reconstructed to be \ca 0.73 \gammaconteggi.\\
\\ 

\begin{table}[]
\begin{center}
\caption{Gamma-ray flux (\gammaconteggi) in the underground Hall A of LNGS.}
\begin{tabular}{cc}
\hline\noalign{\smallskip}
Energy interval&Gamma flux \\
$[$keV$]$ & [\gammaconteggi]\\
\noalign{\smallskip}\hline\noalign{\smallskip}
0 - 500 &0.51 \\
500 - 1000 &0.12 \\
1000 - 2000 &0.08\\
2000 - 3000 &0.015 \\
\noalign{\smallskip}\hline
\end{tabular}
\label{tab:gammaflux}
\end{center}
\end{table}

\subsection{Neutrons}\label{sec:SOURCES_neutrons}

In a deep site the neutron flux below \ca10 MeV  is mainly due to spontaneous fission (the most important source being \udt) and ($\alpha$,n) processes due to interactions of $\alpha$'s from natural emitters with light target nuclei in the rock. Neutrons with energies above \ca10~MeV are produced by nuclear reactions induced by cosmic ray muons in the rock and their flux can extend up to a few GeV. Results obtained by neutron flux measurements at LNGS~\cite{Arneodo,Belli} are reported in Tab.~\ref{tab:nLNGS}. Simulations for both radioactivity induced and muon induced neutrons are found in literature~\cite{Wul2004_1,Wul2004_2,Hime2006}.  In the case of radioactivity induced neutrons the comparison with experimental results is rather good, while in the case of muon induced neutrons this comparison is not possible since this component dominates the environmental neutron flux only above \ca10 MeV, a region where poor experimental data is available. 
This flux can be however computed on the basis of Monte Carlo simulations. Existing simulations performed to evaluate the muon induced neutrons at LNGS~\cite{Wul2004_1,Hime2006}  yield an integral flux above 10 MeV  that is three orders of magnitude lower than the radioactivity induced one, that amounts to \ca $4\cdot10^{-6}$ \neuconteggi below 10 MeV.
Also muon induced cascades or showers in experimental materials (especially high A materials for gamma-ray shielding such as lead) can be an important source of hard neutrons.

Neutrons are absorbed after moderation, this step being generally obtained by using several cms layer of hydrogen rich compounds (e.g. water and polyethylene). Neutron absorption is instead based on high neutron capture cross section isotopes (B and Li are among the most used).

\begin{table}[]
\begin{center}
\caption{Neutron flux measurements at the Gran Sasso laboratory~\cite{Arneodo,Belli}.}
\begin{tabular}{ccc}
\hline\noalign{\smallskip}
Energy interval	&	\multicolumn{2}{c}{Neutron flux[10$^{-6}$ \neuconteggi]} \\
$[$MeV$]$       & 	Ref.~\cite{Arneodo}        & Ref.~\cite{Belli}\\
\noalign{\smallskip}\hline\noalign{\smallskip}
10$^{-3}$ - 2.5 & 	-       		   & 0.54 $\pm$ 0.01\\
1 - 2.5 	& 	0.14 $\pm$ 0.12		   & -              \\
2.5 - 5         & 	0.13 $\pm$ 0.04		   & 0.27 $\pm$ 0.14\\
5 - 10 	        & 	0.15 $\pm$ 0.04		   & 0.05 $\pm$ 0.01\\
10 - 15         & 	(0.4 $\pm$ 0.4) 10$^{-3}$  & (0.6 $\pm$ 0.2) 10$^{-3}$\\
15 - 25         & 	                           & (0.5 $\pm$ 0.3) 10$^{-6}$\\
\noalign{\smallskip}\hline
\end{tabular}
\label{tab:nLNGS}
\end{center}
\end{table}

\subsection{Muons} \label{sec:SOURCES_muons}

The cosmic muon flux at LNGS is reduced by about six orders of magnitude with respect to the outside flux by the 3600 m.w.e. shield, provided by the mountain itself ~\cite{LVD}. The average muon energy in the underground site is  270 GeV~\cite{macro1} with an integrated flux of \ca (3.2 \pom 0.2)$\cdot 10^{-8}$ \muconteggi~\cite{macro2} and average zenith angle $\langle\theta\rangle\sim 35^\circ$. The distribution of the azimuth angle $\phi$ follows the profile of the mountain. The differential flux and zenith angle dependence have been measured by  LVD~\cite{LVD} and MACRO~\cite{macro3}. \\
Generally, direct interactions of muons in the detector are not particularly dangerous. A minimum ionizing particle traversing a CUORE crystal would cause an energy deposition of a few tens of MeV, which is well above the Q-value of \BBz. Moreover the contribution to the background of direct muon interactions as well as of the muon generated electromagnetic showers, can be reduced by means of an anti-coincidence cut between different detectors. On the other hand,  gamma and neutron background arising from muon  interactions in the setup are very dangerous. The presence of high density and high A shields makes the situation even worse (the neutron yield is proportional to $A^{0.8}$~\cite{vitali}) even if simulations of muon produced neutrons in such materials appear to be subject to a large uncertainty. 

Muons, being deep penetrating particles, cannot be removed by a shield. The only way to get rid of this background, in those experiments in which it could be dangerous, is by the use of a muon veto to reject any muon induced interaction.

\section{Geant4 simulation}\label{sec:GEANT4}

Monte Carlo simulations play a primary role in background identification and therefore, in its reduction. Understanding the background event rate in an experiment relies on accurate Monte Carlo modelling of the experimental apparatus and Physics processes leading to all types of background from all possible sources.

In this work, we present the results obtained with the version 4.8.2 of the GEANT4~\cite{geant} simulation tool. GEANT4 is widely used among the particle physics community and its reliability has been established in a number of validation studies.

The LBE (Low Background Experiments) physics list has been used. It is suitable to handle both low energy interactions and hadronic interactions. It contains the Standard and Low Energy electromagnetic processes as well as a complete description of hadronic interactions at low and high energy. The High Precision (HP) neutron models are used to model capture, elastic scattering and inelastic scattering of neutrons from thermal energies up to 20 MeV. 
Gamma, electron and positron production thresholds for secondaries have been set to 5 cm for lead and to 500$\mu$m for copper and iron.

\begin{figure}[t]
\begin{center}
\includegraphics[width=0.5\linewidth, angle=90]{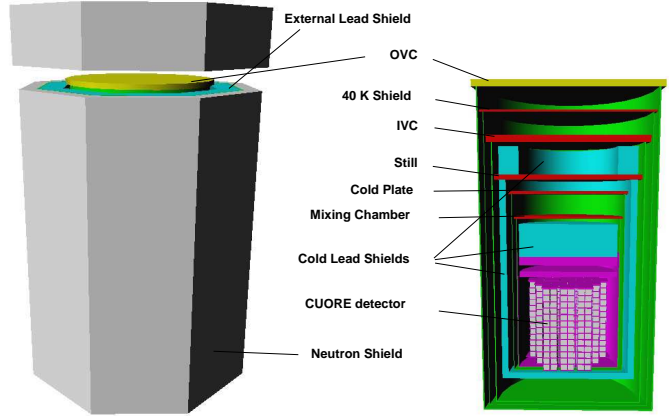}
\end{center}
\caption{3D view of the geometry implemented in the simulation: external octagonal PET shield (left) and internal shields and detector (right).}
\label{fig:CUORE_SimGeom}
\end{figure}

The CUORE geometry used in the simulation is shown in Fig.~\ref{fig:CUORE_SimGeom}. The setup of the external shield has been simplified with respect to the real CUORE design, described in section~\ref{sec:DetStruct}. It consists of an octagonal structure surrounding the cryostat, with a 20 cm neutron shield as the more external layer and a 25 cm octagonal Pb shield inside. 
The 988 detectors are arranged in 19 vertical towers, each tower consisting of 13 layers of four cubes each, as in the real CUORE design. The copper frame structure used to assemble the detector towers and the wire and source trays are also implemented. The small detector details (pins, thermistors, gold wires, teflon spacers, glue spots, etc.) are not implemented, due to their negligible influence on the results concerning environmental contribution to the experimental background (a few tens of grams total mass against  the hundreds tons mass of the shielding). 

The output of the Monte Carlo simulation is converted in detector output by means of an additional code, G2TAS ~\cite{Bucci09}.  In order to take into account  the features of the detector response, namely the slow pulse development and the dead-time, two events depositing energy in the same detector are considered as a single event if the time interval between them is less than an integration time of 10 ms. 
A decay rate can be specified  for each source, than all pile-up events (two events in the same bolometer within the acquisition window of 4 s) are rejected. Coincidence patterns between the detectors (based on event multiplicity, evaluated both in the entire detector array and for the nearest neighbors) are invoked when computing spectra and scatter plots, as it is for experimental data. An energy threshold of 50 keV is assumed. The detector resolution is taken into account introducing a gaussian smearing with 5 keV FWHM.
The ROI has been chosen as a symmetric region 40 keV wide around the nominal Q-value of the \BBz of \tect ((2527.517 \pom 0.013 keV). In this region no gamma photoelectric peaks are expected (it has been chosen above the \coss sum gamma line at 2505 keV and below the 2614.7 $^{208}$Tl line).

\section{Simulation results}\label{sec:GEANT4_bkg}
A full Monte Carlo simulation for muon induced background should be performed generating muons in the rock and propagating them and all secondary particles produced in the rock and in the setup materials to the detector. In the present work, we adopted a different approach:
\begin{itemize}
\item Muons have been generated and propagated from the cavern boundary;
\item Neutrons induced by muon interaction in the rock have been separately generated and propagated also from the cavern boundary;
\item  High energy gamma rays induced by muons in the rock have not been simulated since the evaluated flux is much smaller than the one due to rock radioactivity ~\cite{07Pandola}. 
\end{itemize}
The chosen approach leads to an overestimate of the background due to the loss of correlation between neutrons (and other secondaries) and the original muons. This can cause a multiple site event to be identified as a single site one, thus reducing the anti-coincidence cut efficiency.

The contribution of the environmental gammas to the background in the ROI has been evaluated by simulating primary gamma events at 2614.7 keV (indeed the dominant gamma contribution to the background  comes from the $^{208} $Tl gamma line, the strongest natural gamma line above the Q-value).
According to the spectrum measured in the Hall A of LNGS (see section~\ref{sec:SOURCES_gammas}) the flux at this energy is 9$\cdot$10$^{-3}$ \gammaconteggi. The primary gamma events have been uniformly generated on a spherical surface of 500 cm radius, surrounding the most external shield. The background value in the ROI is obtained by multiplying the integral of the Monte Carlo spectrum in that region by the normalization factor between the number of generated events and the actual flux of 2615 keV gammas at LNGS. 

The results are reported in the first row of Tab.~\ref{tab:backgroundDBD} where the total, the global anti-coincidence and the near channel anti-coincidence contributions are  shown. No anti-coincidence cut is applied in the first case, while an anti-coincidence cut between all the 988 detectors and between each detector and its 18 closest ones are applied in the second and in the third case, respectively. The probability for a gamma ray generated from outside the external shield to reach the detector is very low and, due to multiple Compton interactions, the energy released is low and below the \BBz Q-value. No event has been recorded in the Monte Carlo in the ROI, so only an upper limit has been obtained by considering Poisson statistics, well below the CUORE background goal.

\begin{figure}[t]
\begin{center}
\includegraphics[width=0.7\linewidth]{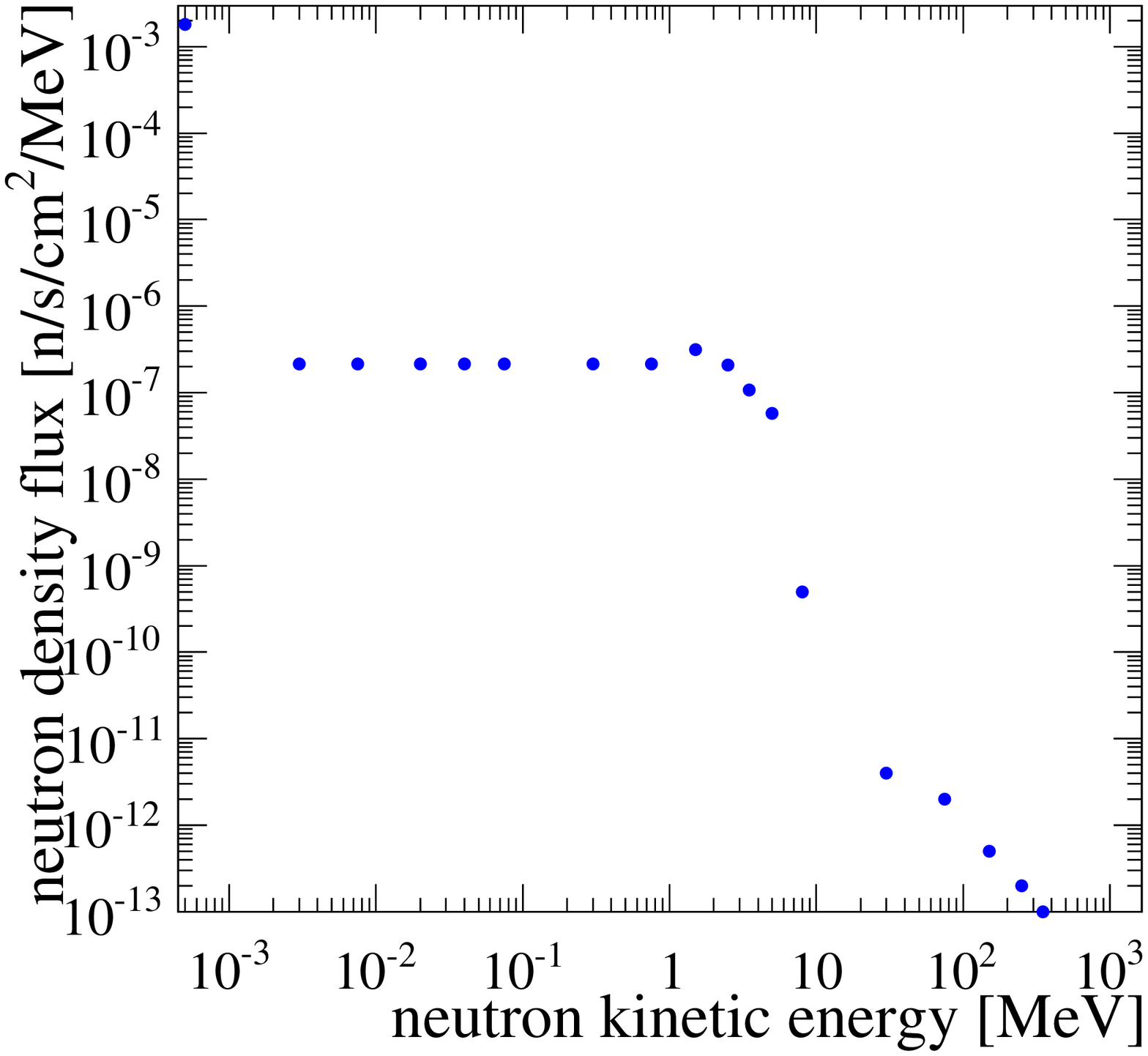}
\end{center}
\caption{Simulated neutron flux density as a function of the neutron kinetic energy. Values are taken accordingly to experimental data ~\cite{Arneodo,Belli} (below 10 MeV) and to existent simulations of muon induced neutrons at LNGS ~\cite{Wul2004_1,Hime2006} (above 10 MeV).}
\label{Fig:nsimulation}
\end{figure}

\begin{figure}[t]
\begin{center}
\includegraphics[width=0.45\linewidth]{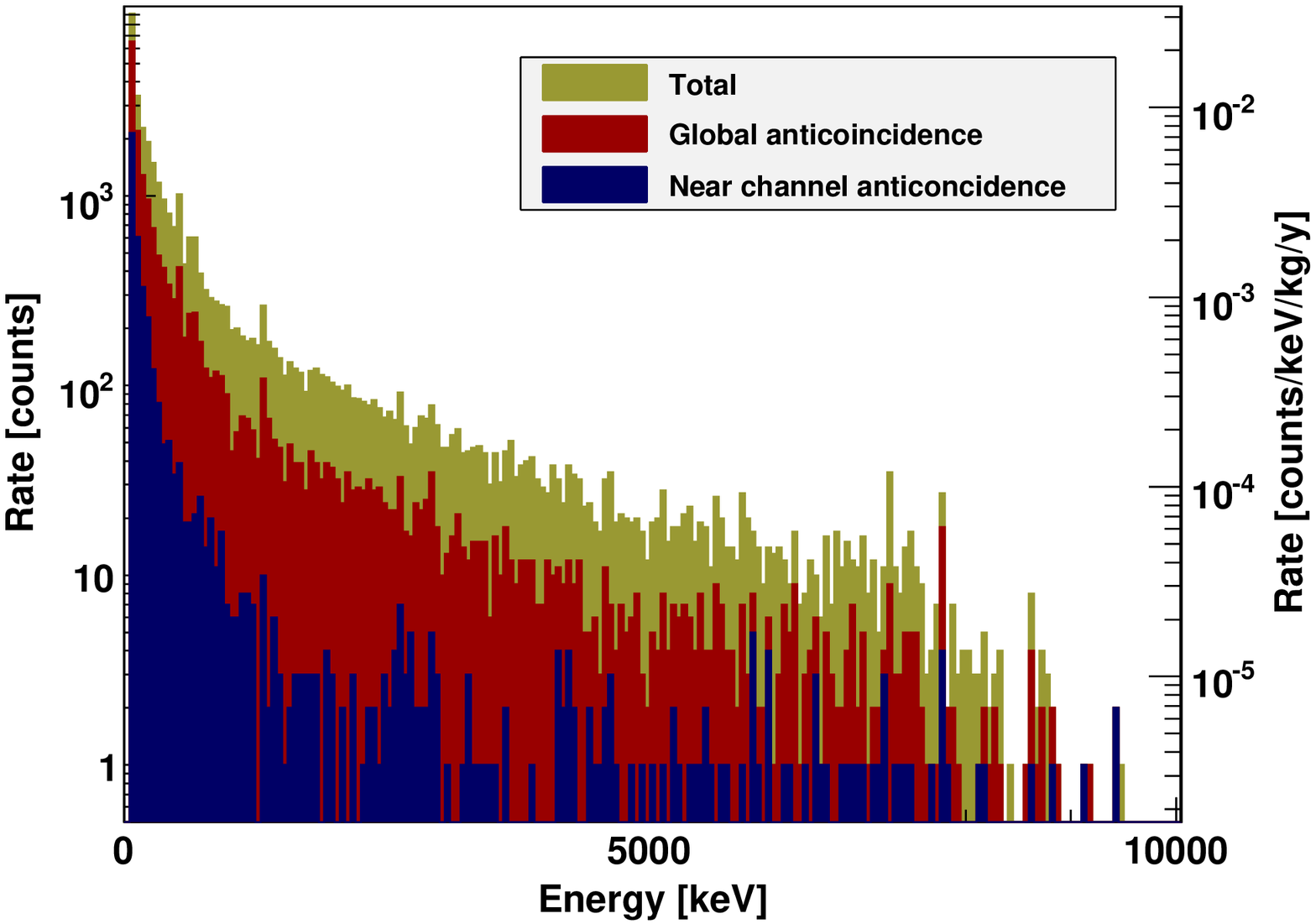}
\hfill
\includegraphics[width=0.45\linewidth]{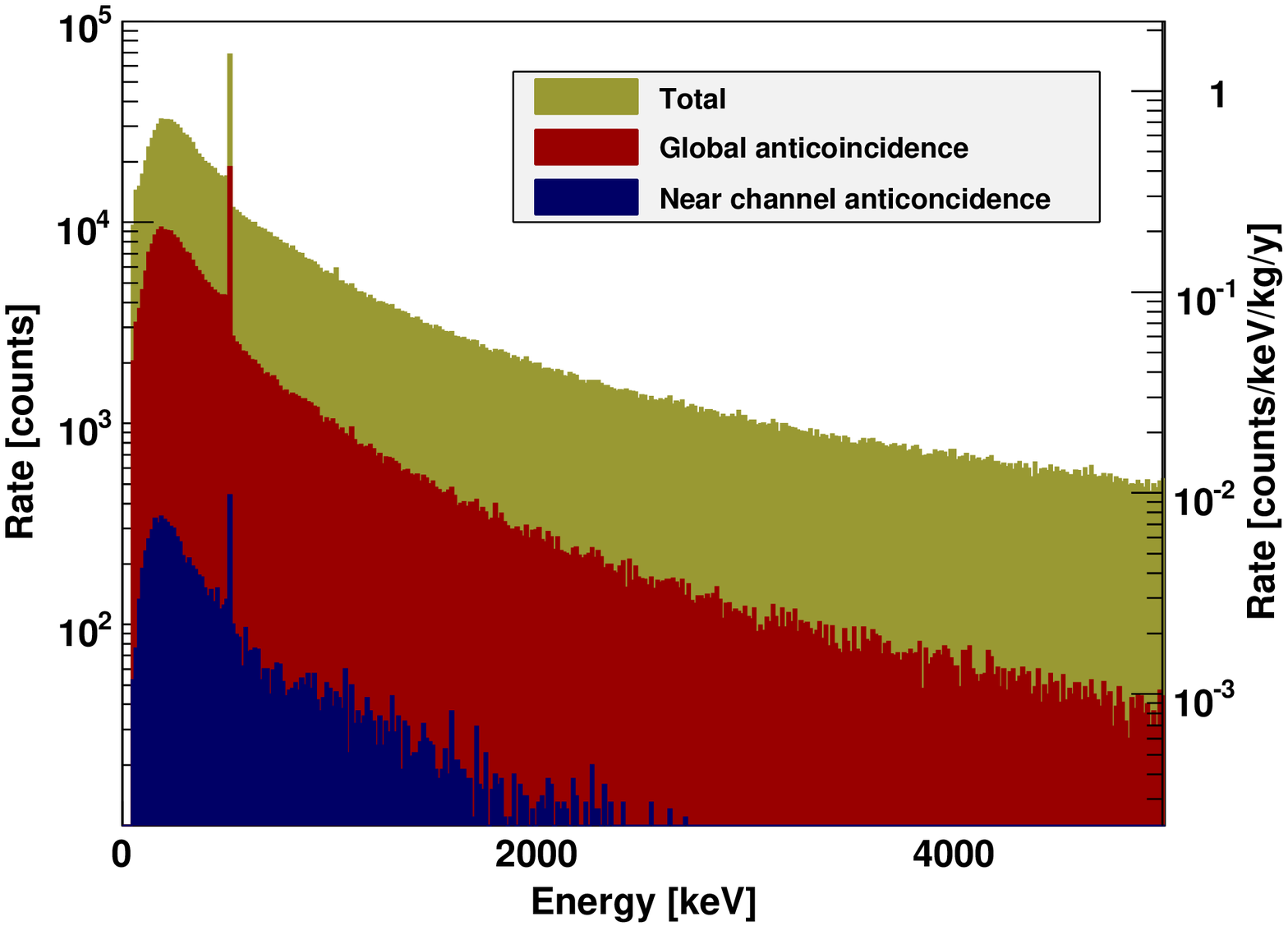}
\end{center}
\caption{Comparison of the Monte Carlo spectra obtained by simulating neutron flux (left) and muon flux (right) at LNGS. In each plot the spectrum obtained without any anti-coincidence cut (green) is shown together with the ones obtained by applying the global anti-coincidence (blue) and the near channel anti-coincidence (red) cut respectively. On the right side of each plot the axis reporting the expected \conteggi in the CUORE detector is shown.}
\label{Fig:neutroni-muoni}
\end{figure}

The neutron spectrum has been simulated according to Fig.~\ref{Fig:nsimulation}. The energy interval between 25 meV (thermal neutrons) and 1.5 GeV has been divided in 22 logarithmic energy bins. Flux values are taken according to existing experimental data~\cite{Arneodo,Belli} below 10 MeV and to simulations of muon induced neutrons at LNGS~\cite{Wul2004_1,Hime2006} above 10 MeV. In order to be conservative, for each energy bin the highest flux among different measurements or different simulations has been used. The total neutron flux, thus extracted, is found to be: $(3.7 \pm 0.2)\cdot10^{-6}$ \neuconteggi. 
The results are summarized in the second row  of Tab.~\ref{tab:backgroundDBD}. Even the total background is well below the CUORE requirement. Furthermore, a reduction factor greater than 1 order of magnitude could be obtained  by applying a global anti-coincidence cut  and a reduction factor of $\sim$5  by applying a near channel anti-coincidence cut. In Fig.~\ref{Fig:neutroni-muoni} a comparison between the neutron background spectrum before and after the global and near channel anti-coincidence cuts is shown. Simulations performed splitting the neutron spectrum in the energy bins of Tab.~\ref{tab:nLNGS} show that the largest contribution is due to neutrons with energies up to 10 MeV, i.e. radioactivity induced neutrons.

\begin{figure}[t]
\begin{center}
\includegraphics[width=0.7\linewidth]{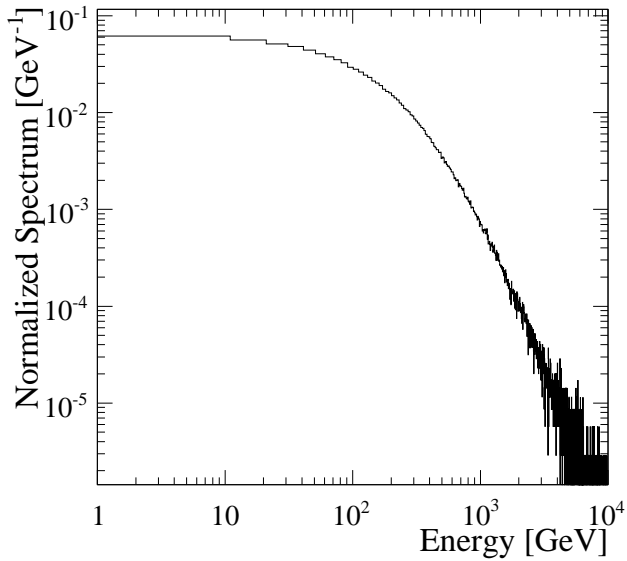}
\end{center}
\caption{Simulated muon energy spectrum.}
\label{fig:MuonSimulSpe}
\end{figure}

Muon induced background was modelled by generating muons on a spherical surface of 500 cm radius (similar to gamma rays and neutrons) and transporting these muons along with all secondary particles produced in muon interaction in the detector materials, through the whole setup. All energy depositions in the crystals were recorded. The muon spectrum  angular distribution is parameterized on the MACRO data (see section~\ref{sec:SOURCES_muons}). The energy spectrum is generated at the ground level (GL) according to ~\cite{Amsler:2008zzb}, that in the high energy limit ($E_{GL}>1$ TeV) can be approximated to ~\cite{macro1}:

\begin{equation}
\frac{dN}{dE_{GL}\cdot d\Omega} \propto \frac{0.14\cdot E^{-3.7}_{GL}}{s\; cm^2\; sr\; GeV} \;\Big(\frac{1}{1+\alpha\;E_{GL}\;\cos\theta}+\frac{0.054}{1+\beta\;E_{GL}\;\cos\theta}\Big)
\end{equation} 

\noindent where $\alpha$=1.1/115 GeV and $\beta$=1.1/850 GeV. It is then extrapolated  underground using the equation ~\cite{Amsler:2008zzb}:

\begin{equation}
E_{U} = (E_{GL}+\epsilon)\cdot E^{-bX} - \epsilon
\end{equation} 

\noindent where b=0.4$\cdot 10^{-6}$ g$^{-1}$ cm$^{-2}$, $\epsilon$=540.0 GeV and $X$ is the thickness of the rock traversed by the muon. $X$ is computed for each given direction  taking into account amount and shape of the overburden mountain according to MACRO data. The simulated muon energy spectrum is shown in Fig.~\ref{fig:MuonSimulSpe}. The energy mean value is consistent with the measured one of 270 GeV ~\cite{macro1,cribier97}.

The muon energy has been chosen between 1 GeV and 10 TeV underground, including 99.2\% of the total muon flux. 
The resulting total background from cosmic ray muons (third row of Tab.~\ref{tab:backgroundDBD}) is higher than the CUORE goal. A reduction factor of almost two orders of magnitude is obtained  by applying a global anti-coincidence cut  and a reduction factor of $\sim$10 by applying the near channel anti-coincidence (Fig.~\ref{Fig:neutroni-muoni}). The CUORE background requirement  could thus be reached without the necessity of a muon veto system.

\begin{table}[]
\begin{center}
\caption{Background in the ROI obtained by simulations of the evaluated gamma, neutron and muon fluxes at LNGS. Quoted errors are statistical only. Units are: $counts\cdot10^{-3}/kg\,keV\,yr$. Limits are given at 90\% C.L. }
\begin{tabular}{ccccc}
\hline\noalign{\smallskip}
Source & Statistics  & Total & Anti-coinc. & Anti-coinc. \\
                 & [years]     &  & (Global) & (Near chan.)\\
\noalign{\smallskip}\hline\noalign{\smallskip}
Gamma   & 0.1   & 0.390          &  0.390        &  0.390    \\
Neutron	 &  7.88  & $0.270\pm0.022$ & $(8.56\pm6.06)\times$10$^{-3}$ & $0.0642\pm0.0442$ \\
Muon  & 4.05    & $17.3\pm0.3$ & $0.104\pm0.022$ & $1.850\pm0.049$ \\
\noalign{\smallskip}\hline
\end{tabular}
\label{tab:backgroundDBD}
\end{center}
\end{table}

The errors quoted in Tab.~\ref{tab:backgroundDBD} and for the background results are statistical only. 
The neutron flux above 10 MeV, due to cosmic muon interactions in the rock surrounding the laboratory, has a systematic error of about a factor 2. This value takes into account  both the discrepancies between the Monte Carlo calculations performed with different codes (i.e. Geant4 and Fluka) \cite{05Araujo} and between the calculations and the available neutron yield data  for light targets and for heavy targets like lead \cite{vitali}. Below 10 MeV the systematic uncertainty on the measured neutron flux is evaluated to be 6.5\% for \cite{Belli} and 31.6\% for \cite{Arneodo} and the two measures have discrepancies up to a factor 3.  Also with the conservative approach adopted (see above) the event rate is much below the CUORE target.  An additional source of systematics in the background evaluation is due to neutron propagation and detection in the simulation code, mainly related to the precision of the cross section tables used to model capture, elastic and inelastic scattering. According to \cite{07Pandola} it has been assumed to be  20\%.

The systematic error in the muon flux is 8\% \cite{macro2}. Uncertainties in the muon induced neutron yield, neutron transport and detection must be also taken into account as a source of systematic error together with the uncertainties in the models used in Geant4 to evaluate the electromagnetic component. The former has a total error of 45\%, the latter of 5\% \cite{07Pandola}.

The uncertainties in the environmental gamma flux at LNGS originate from the measurement statistics and from the detector efficiency evaluation, performed with Monte Carlo simulations. The gamma peaks are measured with uncertainties of less then 8\%, while the simulation uncertainties in the Geant4 electromagnetic tracking have been evaluated to be at a few percent level \cite{07Pandola}. Gamma transport, interaction and detection uncertainties in the Monte Carlo code used to extrapolate the CUORE background are an additional source of systematics at a few percent level.

\section{Conclusions}
The contribution of environmental and cosmic radioactivity at LNGS (i.e. neutrons, gammas, and cosmic muons) to the expected background in the CUORE ROI has been evaluated by means of simulations performed with a GEANT4 based code. Even without the use of a muon veto, the chosen shield configuration guarantees that the overall contribution of these sources is less than $<10^{-2} $\conteggi by operating detectors in near channel anti-coincidence, a value that satisfies the CUORE background goal.

\pagebreak

\end{document}